\documentclass[twocolumn,showpacs,amsmath,amssymb]{revtex4}
\usepackage{graphicx}
\usepackage{dcolumn}
\usepackage{bm}
\usepackage{relsize}

\newcommand{\beq}{\begin{equation}}
\newcommand{\eeq}{\end{equation}}

\begin{document}

\title{Spin relaxation of hot electrons in the $\Gamma$-valley of zinc-blende semiconductors}
\author{Yang Song}\email{yangsong@pas.rochester.edu}
\author{Hanan Dery}\altaffiliation[Also at ]{Department of Electrical and Computer Engineering, University of Rochester, Rochester, New-York, 14627}
\affiliation{Department of Physics and Astronomy, University of
Rochester, Rochester, New-York, 14627}
\begin{abstract}
We present a technique to calculate the spin relaxation of hot
electrons during their energy thermalization in the $\Gamma$-valley
of zinc-blende semiconductors. The results of this model match
recent experimental data and they can be used to check the
applicability of spin related boundary conditions across a
semiconductor/ferromagnet junction in reverse bias conditions (spin
injection).
\end{abstract}
\maketitle

Spin relaxation has an important role in semiconductor spintronics
devices \cite{Zutic_RMP04}. Recent measurements by Crooker
\textit{et al.} have shown that the spin polarization across a
reverse bias GaAs/Fe junction drops appreciably already in
moderate bias-voltage levels \cite{Crooker_PRB09}. Saikin \textit{et
al.} have studied the spin relaxation of injected hot electrons for
relatively large bias conditions and across a wide GaAs depletion region
\cite{Saikin_JPCM06}. The authors have included zone edge phonon interactions in a Monte Carlo modeling and simulated the spatial spin distribution of injected hot electrons in the $\Gamma$, $X$ and $L$ valleys of GaAs. These effects were then verified by the electroluminescence of Fe-based spin light emitting diodes \cite{Mallory_PRB06}. Ivchenko \textit{et al.} have studied the polarization relaxation of photo-excited hot excitons during a
cascade of longitudinal optical (LO) phonon emissions \cite{Ivchenko_SPST78}. They have
used a method of invariants to determine the general form of the
density matrix. None of these models nor the conventional spin relaxation description of quasi-thermal electrons
\cite{Dyakonov_JETP33, Optical_Orientation} is suitable for
studying the spin relaxation in moderate bias conditions
\cite{Crooker_PRB09}. In this paper, we fill this gap and describe a
transparent cascade spin relaxation process during the ultra-fast
energy thermalization in the $\Gamma$-valley of zinc-blende
semiconductors. The energy thermalization is governed by emission of
long wavelength LO-phonons \cite{Born_Book}.
The results of this model match the experimental data and they can
be used to evaluate the applicability of the boundary conditions in
modeling spintronics devices.



The Dyakonov-Perel mechanism \cite{Dyakonov_JETP33} is by
large the dominant spin relaxation process during thermalization of
hot electrons. However, the assumptions that are used in
calculating the spin relaxation time of quasi-thermal electrons \cite{Optical_Orientation} are not valid for hot electrons. The broken assumptions are that (I) the density matrix of hot electrons is far from equilibrium and often is anisotropic, (II) the scattering is not elastic, and (III) the spin precession of hot electrons due to the intrinsic magnetic field is non negligible even within the momentum relaxation time (along certain high symmetry directions).

The spin behavior of hot electrons should be studied after each momentum
scattering rather than in the asymptotic limit (after many scattering events). The reason is that hot electrons reach the bottom of the conduction band after a few LO-phonon emissions and altogether in about a picosecond. We begin by writing the intrinsic spin dynamics in
the $\Gamma$-valley of a zinc-blende semiconductor \cite{Dyakonov_JETP33},
\begin{eqnarray}
\frac {d\mathbf{ S}}{dt} &=& \boldsymbol{ \Omega }\times {\mathbf{
S}}\label{eq:DP}\,, \\ \Omega_j(\mathbf {k}) &=& \alpha \hbar ^{2}
(2m_{sc}^{3} E_{g} )^{-1/2} k_{j} (k_{\ell}^{2} -k_{m}^{2} ) \,. \label{eq:k3}
\end{eqnarray}
The intrinsic Larmor frequency vector, $\boldsymbol{\Omega}$, has a cubic dependence in the wavevector components \cite{Dresselhaus_PR55}. 
The $\{j,\ell,m\}$ subscripts denote any cyclic permutation of
$\{x,y,z\}$. $E_{g}$ is the energy-gap, $m_{sc}$ is the
electron's effective mass, and $\alpha$ is a dimensionless parameter
that relates to the strength of the spin-orbit coupling in the
semiconductor.

We assume that the initial energy of a hot electron is well
above the $\Gamma$ point but still below the energy of the other
valleys in the conduction band. In this case the energy
thermalization is governed by emission of long wavelength LO-phonons via the Fr\"{o}hlich interaction \cite{Born_Book}. After
each of the ultra-fast momentum scattering events the direction and
magnitude of $\boldsymbol{ \Omega }$ are randomized. The probability
distribution that an electron with a wavevector $\mathbf{k}$ will be
scattered into wavevector $\mathbf{k}'$ is,
\begin{eqnarray}
I(\mathbf{k} \rightarrow \mathbf{k}') &=&  \frac{\mathcal{B}(k) \delta
(\varepsilon(k') + \varepsilon_{LO}- \varepsilon(k))} { k'^2
+k^2 -2k'k\cos (\theta_{k',k} )} \label{eq:Ik}\,.
\end{eqnarray}
The denominator describes the phonon wavevector dependence of the
Fr\"{o}hlich interaction,
$|$$<$$\mathbf{k}'$$|$$\mathcal{H}_F$$|$$\mathbf{k}$$>$$|^2$
$\propto|\mathbf{k}'-\mathbf{k}|^{-2}$. We have neglected the
weak dispersion relation between the energy and wavevector in the
LO-phonon branch, $\varepsilon_{LO}(\mathbf{k}'-\mathbf{k})
\approx \varepsilon_{LO}$. $\mathcal{B}(k)$ is a normalizing
factor and it is proportional to the momentum scattering time from
state $\mathbf{k}$,
\begin{eqnarray}
\frac{1}{\tau_{LO}(k)} \! \propto \! \frac{1}{\mathcal{B}(k)} \!= \!
\frac{k}{ 8\pi^2 \varepsilon(k) }
 \ln \! \left(\frac{1+\sqrt{1-\varepsilon_{LO} /\varepsilon({k}) }}{1-\sqrt{1-\varepsilon_{LO} /\varepsilon(k) } } \, \right) \,.
\label{eq:tauLO}
\end{eqnarray}
The momentum scattering events are represented by a homogeneous
Poisson process where the spin distribution of a hot electron
immediately before its $n^{th}$ scattering event is denoted by,
\begin{eqnarray}
 \tilde{\mathbf{S}}_{n-1}(\mathbf{k})= \frac{1}{\tau _{LO} (k)} \int
_{0}^{\infty }\!\! dt\,e^{-t/\tau _{LO} (k)}
\mathbf{S}_{n-1} (\mathbf{k},t)\,, \label{eq:spin_dist_1}
\end{eqnarray}
where $\mathbf{S}_{n-1} (\mathbf{k},t)$ is the solution of
Eq.~(\ref{eq:DP}) with an initial condition
$\mathbf{S}_{n-1}(\mathbf{k},0)$ which denotes the spin distribution
of an injected electron immediately after its $(n-1)^{th}$
scattering event. Using the probability distribution in
Eq.~(\ref{eq:Ik}), the spin distribution immediately after the
$n^{th}$ scattering event is denoted by,
\begin{eqnarray}
\mathbf{S}_{n}(\mathbf{k}',0)= \frac{1}{(2\pi)^3}\int \! d^3k \,
\tilde{\mathbf{S}}_{n-1} (\mathbf{k})I(\mathbf{k} \rightarrow
\mathbf{k}')\,.
\end{eqnarray}
This cascade process is complete by defining the initial spin
distribution of the hot electrons. Due to the importance of spin
relaxation in spintronics devices, we consider spin injection from a
ferromagnetic contact into a semiconductor. Thus, the initial spin
distribution is governed by the difference of the spin dependent
transmission coefficients,
\begin{eqnarray}
\mathbf{S}_{0}(\mathbf{k},t=0) \propto |t_{\uparrow}(
\mathbf{k},V)|^{2} - |t_{\downarrow}(
\mathbf{k},V)|^{2},\label{eq:injected_spin}
\end{eqnarray}
where $V$ denotes the reverse voltage bias across the
semiconductor/ferromagnet junction. We use a simple  effective mass
model ($\varepsilon(k)=\hbar^2 k^2/2m_{sc}$) to calculate the spin
dependent transmission coefficients due to tunneling across a thin
triangular-like Schottky barrier \cite{Osipov_PRB04}. The results of
the following discussion are not qualitatively affected by the
choice of initial distribution.

The temperature dependence of this cascade process is manifested
only by a slight modification of the intrinsic Larmor frequency vector (via a
change in band-gap energy). However, one should recall that
LO-phonon absorptions may also occur during the thermalization. The
probability of an absorption process is smaller than the probability
of an emission process by $\text{exp}\{-\varepsilon_{LO}/k_BT\}$. Thus, in
typical room temperature III-V semiconductors the absorption
processes may somewhat enhance the spin relaxation of hot-electrons.
In this paper, we ignore absorption processes by limiting our
discussion to low temperatures. One can construct, however, a more
general probabilistic model that incorporates both emission and
absorption processes in the iterative procedure.

\begin{figure}[htp!]
\includegraphics[width=7cm]{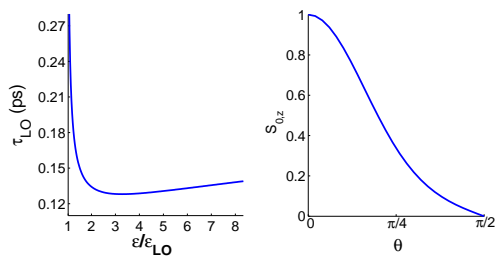}
\caption[LO_DP] { (a) Momentum scattering time in GaAs versus energy (in
$\varepsilon_{LO}$=36~meV units). (b) A typical normalized initial
spin distribution of injected electrons from the Fermi level of Fe
into GaAs versus the polar angle (with respect to the interface
normal). The injected spin direction is aligned with the spin
direction of majority electrons in Fe.}
\label{fig:tau_and_initial}
\end{figure}

\begin{figure}[tp!]
\includegraphics[width=8cm]{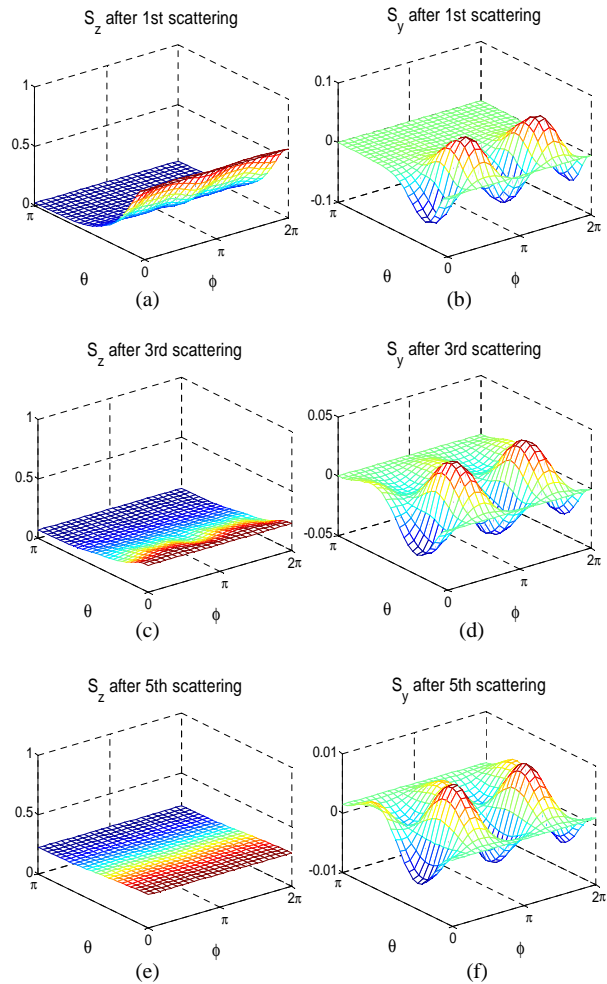}
\caption[LO_DP]
{(color online) (a) Spin distribution over electrons' momentum
polar and azimuth angles after the 1$st$, 3$rd$ and 5$th$ successive
LO-phonon emissions. The energy of the injected electrons is 0.2~eV
above the conduction band $\Gamma$ point and their spin distribution
is normalized and points in the $+z$ direction (see
Fig.~\ref{fig:tau_and_initial}b). $S_z$ keeps its dominant role
throughout the energy thermalization process whereas $S_x$ (not
shown) and $S_y$ are of much lower amplitude. After thermalization
the electrons have lose $\sim$50\% of the net injected $S_z$ value
(see text).
}\label{fig:LO_DP}
\end{figure}

We use this cascade spin relaxation procedure to simulate the spin
relaxation of hot electrons in a 10$^{16}$~cm$^{-3}$ n-type GaAs. To
calculate the intrinsic Larmor frequency vector, $\boldsymbol{
\Omega }(\mathbf{k})$, we use $E_g=1.519$~eV, $m_{sc}$=0.067, and
$\alpha=0.07$ \cite{Marushchak_SPSS83}. To calculate $\tau _{LO}(k)$
from Eq.~(\ref{eq:tauLO}), we assume that its minimum is at 130~fs
and that $\varepsilon_{LO}$=36~meV \cite{Conwell_IEEE66}.
Fig.~\ref{fig:tau_and_initial}a shows the energy dependence of the
momentum scattering time. Fig.~\ref{fig:tau_and_initial}b shows a
typical normalized initial spin distribution of injected electrons
from the Fermi level of Fe versus the angle from the
semiconductor/ferromagnet interface normal. The strongest spin
polarization is from electrons  whose motion is along the interface
normal ($\cos{\theta} \rightarrow 1$). This distribution is
independent of the azimuthal angle ($\phi$) that lies in the
semiconducotr/ferromagnet interface plane. The injected spin
direction is aligned with the spin direction of majority electrons
in Fe and we set it as the $S_z$ component \cite{Hanbicki_APL02}.
Fig.~\ref{fig:LO_DP} shows the evolution of the spin distribution
after the first, third and fifth successive LO-phonon emissions of
hot electrons whose initial energy is 0.2~eV above the semiconductor
conduction band minima. Assuming that these electrons are injected
from the iron's Fermi level, then at 4K this scenario describes a
GaAs/Fe junction at $-$0.2~V reverse bias-voltage (or $-$0.3~V at
300K). It takes five successive LO-phonon emissions for electrons to
reach the vicinity of the $\Gamma$ point. Fig.~\ref{fig:LO_DP}a
shows that after the first LO-phonon emission, the main contribution
to $S_z$ is still from electrons whose motion is along the interface
normal. This can be understood due to the forward scattering nature
of the Fr\"{o}hlich interaction (see the denominator of
Eq.~(\ref{eq:Ik})). Thus, the initial distribution is not totally
randomized in wavevector space after the first LO-phonon emission.
This is not the case after five LO-phonon emissions where we see
that the contribution to the net spin is nearly isotropic in the
electron momentum. At these injection levels the spin information is
largely kept when the electrons reach the bottom of the conduction
band, and  $S_z$ is much larger than the $S_x$ (not shown) and $S_y$
(shown) components.

The important figure of merit is to determine the net spin
of the injected electrons after their ultra-fast energy
thermalization. Using the effective mass approximation, the net spin
after $n$ LO-phonon emissions is given by $k_n^2 dk_n \int d \phi
\int d \theta \sin \theta \, S_{n,i}(\mathbf{k}_n,0)$, where $i$
enumerates the spin components. This writing assumes that initially
the electrons' energy is concentrated in a thin energy shell
between $\varepsilon_0$ and $\varepsilon_0+d\varepsilon_0$. Due to
the energy conservation of the Fr\"{o}hlich interaction
(Eq.~(\ref{eq:Ik})), the ratio between the $k_n^2 dk_n$ and $k_0^2
dk_0$ prefactors simply yields $k_n/k_0$. Thus, the fraction of the
net injected spin that is left after the thermalization process is
denoted by,
\begin{eqnarray}
R(\varepsilon_0) = \left(\!\sqrt{\frac{\varepsilon_0}{\varepsilon_0 - \, \scriptstyle{N \cdot} \varepsilon_{LO}}}\,\right) \cdot \frac{\int \! d \phi \!\int\! d \theta \sin \theta \, S_{N,z}(\mathbf{k}_N,0)}{\int \!d \phi \!\int\! d \theta \sin \theta \, S_{0,z}(\mathbf{k}_0,0)} \,,
\end{eqnarray}
where $\varepsilon_0$ is the electrons' injected energy above the
semiconductor conduction band $\Gamma$-point and $N$ is the number
of phonon emissions that are needed to reach the bottom of the band.
We have considered only the $z$-direction since the injected net
spin is aligned along this direction.

\begin{figure}[]
\includegraphics[width=5cm]{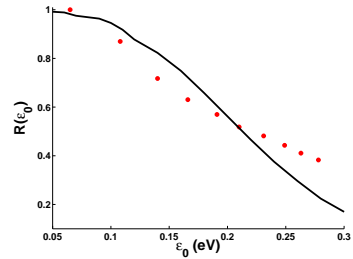}
\caption{The ratio between the final and injected net spin as a
function of the initial injected energy. The dots denote the
measured values by Crooker \textit{et al.} \cite{Crooker_PRB09}.}
\label{fig:Ratio}
\end{figure}

Fig.~\ref{fig:Ratio} shows the ratio between the final and initial
injected net spin. The spin information is largely kept when the
energy of injected electrons is less than 0.2~eV. These results are
also in accordance with recent measured data by Crooker \textit{et
al.} \cite{Crooker_PRB09}. The slower spin decay of the experimental
results is attributed to the fact that the theoretical curve
considers only the most energetic injected electrons (whose spin
relaxation is fastest). In the experiment, however, electrons with
lower energy also contribute to the current (the allowed injection
energy is anywhere between the conduction band edge in the
semiconductor and the Fe Fermi level). The application of this
iterative procedure is limited if the injected electrons experience
strong inter-valley scattering processes \cite{Saikin_JPCM06}. In
GaAs this limit is reached when the injection energy is
$\varepsilon_0$=0.3~eV \cite{Yu_Cardona}. Our model shows that less
than 1/5 of the net injected spin is left during the possible $n=8$
LO-phonon emissions in this limit.

In summary, we have presented a transparent technique to trace the
spin evolution during the relaxation of hot electrons in the
conduction band $\Gamma$-valley of zinc-blende semiconductors. We
have shown that if the voltage drop across a reversed biased GaAs/Fe
junction is moderate then the spin information is largely kept. This
technique can be used to evaluate the applicability of the boundary
conditions in modeling spintronics devices. 

We thank Professor P. Crowell for communicating
the experimental results. This work is supported by AFOSR Contract No. FA9550-09-1-0493 and by
NSF Contract No. ECCS-0824075.

\end{document}